\documentclass[12pt]{article}
\usepackage{geometry}
\usepackage{graphicx}
\usepackage{amssymb,amsmath,amsfonts,amsthm,exscale,latexsym}
\usepackage{natbib}
\usepackage[flushleft]{threeparttable}
\renewcommand{\baselinestretch}{1.5}
\renewcommand{\arraystretch}{1}

\usepackage[linktoc=page,colorlinks=true,linkcolor=blue,citecolor=blue,urlcolor=blue]{hyperref}
\usepackage{url}

\begin{document}

\begin{center}
	{\large \bf Improving estimation efficiency of case-cohort study with interval-censored failure time data}
\end{center}

\begin{center}
Qingning Zhou\textsuperscript{1} and Kin Yau Wong\textsuperscript{2}

\textsuperscript{1}Department of Mathematics and Statistics, University of North Carolina at Charlotte

\textsuperscript{2}Department of Applied Mathematics, The Hong Kong Polytechnic University
\end{center}

\vspace{0.1in}

\begin{abstract}
The case-cohort design is a commonly used cost-effective sampling strategy for large cohort studies, where some covariates are expensive to measure or obtain. In this paper, we consider regression analysis under a case-cohort study with interval-censored failure time data, where the failure time is only known to fall within an interval instead of being exactly observed. A common approach to analyze data from a case-cohort study is the inverse probability weighting approach, where only subjects in the case-cohort sample are used in estimation, and the subjects are weighted based on the probability of inclusion into the case-cohort sample. This approach, though consistent, is generally inefficient as it does not incorporate information outside the case-cohort sample. To improve efficiency, we first develop a sieve maximum weighted likelihood estimator under the Cox model based on the case-cohort sample, and then propose a procedure to update this estimator by using information in the full cohort. We show that the update estimator is consistent, asymptotically normal, and more efficient than the original estimator. The proposed method can flexibly incorporate auxiliary variables to further improve estimation efficiency. We employ a weighted bootstrap procedure for variance estimation. Simulation results indicate that the proposed method works well in practical situations. A real study on diabetes is provided for illustration.

\vspace{.1in}
\end{abstract}

\newpage

\section{Introduction}
The case-cohort design has been widely used as a cost-effective sampling strategy to conduct epidemiological and biomedical studies where the outcomes of interest are times to some rare events, such as HIV infection and the onset of coronary heart disease, and some covariates are expensive to collect or measure. For example, when covariate measurements involve expensive bio-assay, genetic measurements, or labor-intensive chart review, it may be economically infeasible to conduct the measurements for all study subjects. The case-cohort design, in which only the cases and a subcohort of subjects are selected for the expensive covariate measurements, aims to yield more efficient inference under a certain budget constraint. Since its initial proposal by \cite{prentice1986case}, there has been extensive research in the statistical analysis of case-cohort design and its variations. Among others, under the Cox model, \cite{prentice1986case} and \cite{self1988asymptotic} proposed a pseudo-likelihood method for estimation and inference; \cite{chen1999case} developed an estimating equation approach; \cite{marti2011multiple} proposed a multiple imputation method; \cite{scheike2004maximum} and \cite{zeng2014efficient} considered maximum likelihood estimation. Moreover, \cite{kang2009marginal},  \cite{kim2013more} and \cite{kim2018analysis} developed estimating equation approaches for case-cohort design with multiple outcomes. However, most of the existing methods for case-cohort design were developed for right-censored survival data. 

In this paper, we consider the case-cohort design with interval-censored failure time data. Interval censoring occurs when the failure time of interest is only known to fall within an interval rather than being exactly observed. Interval-censored data commonly arise in epidemiological and biomedical studies that involve asymptomatic events, such as HIV infection, the onset of AIDS, and the onset of diabetes. For example, in the Atherosclerosis Risk in Communities (ARIC) study considered in Section 7, the outcome of interest is time to the onset of diabetes. The participants were examined every three years and they may miss some examinations or may not come at the scheduled times. If a participant was observed to have diabetes, it was known only that the onset of diabetes occurred between two consecutive visits. Interval censoring complicates the likelihood and poses challenges for estimation and inference.

Research on the case-cohort design with interval-censored data is limited. \cite{li2008weighted} and \cite{li2011relative} studied case-cohort design with grouped survival data and current status data, respectively, which are special cases of interval-censored data. \cite{zhou2017case} considered case-cohort design with general interval-censored data and proposed a sieve maximum weighted likelihood method based on inverse probability weighting (IPW).  \cite{zhou2021semiparametric} studied case-cohort design with multiple interval-censored outcomes and developed an IPW method with weights that incorporate information from multiple events. \cite{du2021regression} considered case-cohort design with informatively interval-censored data, where the monitoring times depend on the failure time, and also employed an IPW method to handle the sampling bias induced by the case-cohort design. It is well known that the IPW method only uses data in the case-cohort sample and is inefficient. We propose an update estimation procedure that improves the efficiency of the IPW estimator in \cite{zhou2017case} by using information from the full cohort. 

The main idea of the update approach is to find an (asymptotically) mean-zero statistic that is potentially correlated with the original unbiased estimator and then construct an update estimator as the optimal linear combination of the original estimator with the mean-zero statistic. It can be shown that the update estimator is still unbiased and at least as efficient as the original estimator, and is more efficient if the mean-zero statistic is correlated with the original estimator. This update approach has been employed under different settings involving incomplete or imprecise data \citep{chen2000unified,chen2002cox,wang2015semiparametric,yan2017improving,yang2020combining,tong2020augmented,yin2022cost}. In this paper, we propose an update estimation procedure to improve the efficiency of the IPW estimator given by \cite{zhou2017case} for regression analysis of case-cohort study with interval-censored failure time data. Specifically, we assume a working regression model of the failure time given the cheap covariates and auxiliary variables if available. We then fit the working model to the case-cohort sample and the full cohort data, respectively, to obtain two estimators of the same mean. We thus can find a mean-zero statistic by taking difference of the two estimators and construct an update estimator as the optimal linear combination of the IPW estimator with the mean-zero statistic.

The remainder of this paper is organized as follows. In Section 2, we describe the design, data structure and model assumptions. In Section 3, we review the IPW estimator given by \cite{zhou2017case}. In Section 4, we propose an update estimation procedure to improve the IPW estimator using information from the full cohort. The asymptotic properties of the update estimator are established in Section 5. Simulation studies and an illustrative example are given in Sections 6 and 7, respectively. We conclude in Section 8 with some discussions on extensions or directions for future research.

\section{Design, Data and Model}
Let $T$ be the failure time, $X$ be a vector of expensive covariates, and $Z$ be a vector of cheap covariates. Assume that $T$ follows the Cox model with the conditional cumulative hazard function given by
\begin{equation}\label{model}
	\Lambda(t\,|\,X,Z)=\Lambda(t) \exp(\beta^{\text T} X+\gamma^{\text T} Z),
\end{equation}
where $\vartheta\equiv(\beta^{\text T},\gamma^{\text T})^{\text T}$ is a $d$-vector of regression parameters and $\Lambda$ is the unspecified cumulative baseline hazard function. Also, let $X^*$ denote a vector of cheap auxiliary variables that could be available in practice and informative to the expensive covariate $X$. Assume that $T$ and $X^*$ are independent given $X$ and $Z$.

Suppose that we observe interval-censored failure time data. Let $U_1,\ldots,U_K$ denote the random examination times such that $0=U_0<U_1<\cdots<U_K<U_{K+1}=\infty$, where $K$ is a random positive integer. Also, define $\Delta_k=I(U_{k-1}< T \leq U_k)$ for $k=1,\ldots,K+1$, where $I(\cdot)$ is the indicator function. Then the interval-censored failure time data consists of $\{K,\,U_1,\ldots,U_K, \, \Delta_1,\ldots,\Delta_K\}.$ We assume that the examination times are conditionally independent of the failure time given the covariates. 

We consider a two-phase (generalized) case-cohort design based on a full cohort of size $n$. At Phase I, we observe the interval-censored failure time, the cheap covariates, and the auxiliary variables for all cohort members, denoted by $\{K_i,\,U_{i1},\ldots,U_{iK_i}, \, \Delta_{i1},\ldots,\Delta_{iK_i},\,Z_i,\,X_i^*\}$ for $i=1,\ldots,n$. At Phase II, we first select a subcohort via independent Bernoulli sampling with a known success probability $q_s\in(0,1]$, and then select a subset of cases (i.e., subjects with $\sum_{k=1}^{K_i} \Delta_{ik} = 1$) outside the subcohort also by Bernoulli sampling with a known success probability $q_c\in(0,1]$. Note that if $q_c=1$, then all cases are selected, and this is called a case-cohort design; if $q_c\in (0,1)$, then this is usually referred to as a generalized case-cohort design. Let $\eta_i$ indicate whether the $i$-th subject is selected into the subcohort and $\zeta_i$ indicate whether the $i$-th subject is a selected case. Note that if $\zeta_i=1$, then $\eta_i=0$ and $\sum_{k=1}^{K_i} \Delta_{ik} = 1$. The expensive covariates are observed for subjects in the subcohort and for the selected cases outside the subcohort, that is, for the $i$-th subject, $X_i$ is observed if $\eta_i=1$ or $\zeta_i=1$. Let $\xi_i$ indicate whether $X_i$ is observed. Then the observed data can be represented by
$$
O_i^\xi=\{K_i,\,U_{i1},\ldots,U_{iK_i}, \, \Delta_{i1},\ldots,\Delta_{iK_i},\,Z_i,\,X_i^*,\,\xi_i X_i,\,\xi_i\},\quad i=1,\ldots,n.
$$

\section{Sieve Maximum Weighted Likelihood Estimator}
Let $\theta=(\vartheta,\Lambda)$. For estimation, we consider the weighted log-likelihood function
\begin{equation*}\label{wlik}
	\begin{split}
		l_n^w(\theta)= &\sum_{i=1}^n w_i \Bigg(\sum_{k=1}^{K_i+1} \Delta_{ik}  \log\Big[\exp\left\{-\Lambda(U_{i,k-1})\exp(\beta^{\text T} X_i+\gamma^{\text T} Z_i)\right\} \\
		&\quad\quad - \exp\left\{-\Lambda(U_{ik})\exp(\beta^{\text T} X_i+\gamma^{\text T} Z_i)\right\} \Big]   \Bigg),
	\end{split}
\end{equation*}
where the weight $w_i$ is defined as
$$ 
w_i = \frac{\xi_i}{\pi_q(\Delta_i)} = \frac{\xi_i}{\left(1-\sum_{k=1}^{K_i} \Delta_{ik}\right) q_s + \left(\sum_{k=1}^{K_i} \Delta_{ik}\right) \{q_s+(1-q_s)q_c\} }.
$$
Here we assume that $q_s$ and $q_c$ are known for simplicity; our approach still works if they are unknown and replaced by consistent estimators.

To estimate the unknown function $\Lambda$, \cite{zhou2017case} proposed a sieve method based on Bernstein polynomials. Specifically, the sieve space is defined as 
\begin{equation}\label{sievesp}
	\Theta_n \, = \mathcal{B} \otimes \mathcal{M}_n \, ,
\end{equation}
where $\mathcal{B}$ is a compact set in $R^d$ and 
$$
\mathcal{M}_n = \left\{ \Lambda_{n}(t) = \sum_{k=0}^{m}\phi_{j}
B_j(t,m,\sigma,\tau) : \,\,\,\phi_{m} \geq \ldots \geq \phi_{1} \geq \phi_{0} \geq 0, \,\,
\sum_{k=0}^{m}|\phi_{j}|\leq M_n\right\} 
$$
with $B_j(t,m,\sigma,\tau)$ being Bernstein basis polynomials of degree $m$, i.e.,
$$
B_j(t,m,\sigma,\tau) \,=\, \binom{m}{j}\left(\frac{t-\sigma}{\tau-\sigma}\right)^j 
\left( 1 - \frac{t-\sigma}{\tau-\sigma}\right)^{m-j} \, ,\,\,\, k=0,\ldots,m.
$$
Here $\sigma$ and $\tau$ are the lower and upper bounds of the examination times, respectively, with $0<\sigma<\tau<\infty$. The sieve maximum weighted likelihood estimator $\hat{\theta}_n=(\hat{\vartheta}_{n},\hat{\Lambda}_{n})$ is defined as the value of $\theta$ that maximizes the weighted log-likelihood function $l_n^w$ over $\Theta_n$.

It is well known that the inverse probability weighted (IPW) estimator is inefficient. To improve estimation efficiency, we consider an update approach that utilizes the available information in the full cohort by fitting a working model relating the cheap covariates or auxiliary variables to the failure time. It can be shown that the update estimator is guaranteed to be at least asymptotically as efficient as the original IPW estimator.

\section{Proposed Update Estimator}

Consider a working Cox model for $T$ given $X^*$ and $Z$ with the conditional cumulative hazard function 
\begin{equation}\label{model_work}
	\Lambda^*(t\,|\,X^*,Z)=\Lambda^*(t) \exp(\beta^{* \text T} X^*+\gamma^{* \text T} Z),
\end{equation}
where $\vartheta^*\equiv(\beta^{* \text T},\gamma^{* \text T})^{\text T}$ is a $d^*$-vector of regression parameters and $\Lambda^*$ is the unspecified cumulative baseline hazard function. Note that we can consider a working Cox model given $Z$ only, if $X^*$ is not available.

We first estimate the working Cox model \eqref{model_work} using the Bernstein-polynomial-based sieve method similarly as the above but with covariates $X^*$ and $Z$ instead. Let $\hat{\theta}^*_n=(\hat{\vartheta}^*_n,\hat{\Lambda}^*_n)$ denote the sieve maximum weighted likelihood estimator of $\theta^*=(\vartheta^*,\Lambda^*)$ based on the case-cohort sample. Since $X^*$ and $Z$ are available for all subjects in the cohort, we can also obtain the sieve maximum likelihood estimator of $\theta^*=(\vartheta^*,\Lambda^*)$, denoted by $\bar{\theta}^*_n=(\bar{\vartheta}^*_n,\bar{\Lambda}^*_n)$, based on the full cohort. 
Let $\Sigma=[\Sigma_{11},\Sigma_{12};\Sigma_{21},\Sigma_{22}]$ be the covariance matrix of the limiting distribution of $\big(\sqrt{n}(\hat{\vartheta}_n-\vartheta_0)^{\text T},\sqrt{n}(\hat{\vartheta}^*_n-\bar{\vartheta}^*_n)^{\text T}\big)^{\text T}$, and let $\hat\Sigma=[\hat\Sigma_{11},\hat\Sigma_{12};\hat\Sigma_{21},\hat\Sigma_{22}]$ denote a consistent estimator of $\Sigma$. We define the update estimator of $\vartheta$ as
$$
\bar{\vartheta}_n = \hat{\vartheta}_n-\hat\Sigma_{12}\hat\Sigma_{22}^{-1}(\hat{\vartheta}^*_n-\bar{\vartheta}^*_n).
$$
We can show that the asymptotic covariance matrix of $\sqrt{n}(\hat{\vartheta}_n-\vartheta_0)$ is $\Sigma_{11}$, while the asymptotic covariance matrix of $\sqrt{n}(\bar{\vartheta}_n-\vartheta_0)$ is $\Sigma_{11}-\Sigma_{12}\Sigma_{22}^{-1}\Sigma_{21}$. Thus, $\bar{\vartheta}_n$ is asymptotically at least as efficient as $\hat{\vartheta}_n$.

We propose to estimate the covariance matrix $\Sigma$ using the weighted bootstrap method. Specifically, let $\{u_1,\ldots,u_n\}$ denote $n$ independent realizations of a bounded positive random variable $u$ satisfying $E(u)=\text{var}(u)=1$. 
We use the Exponential distribution with mean one in the simulation study and real data analysis. Define the new weights $w_i^b=u_iw_i$ for $i=1,\ldots,n$. Let $\hat{\theta}_n^b=(\hat{\vartheta}_n^b,\hat{\Lambda}_n^b)$ be the sieve maximum weighted likelihood estimator that maximizes the new weighted log-likelihood function $l_n^{w^b}$ over $\Theta_n$, where $l_n^{w^b}$ is obtained by replacing $w_i$ with $w_i^b$ in $l_n^w$. We generate $B$ samples of $\{u_1,\ldots,u_n\}$ and obtain the corresponding $\hat{\vartheta}_n^b$ as well as $\hat{\vartheta}^{*b}_n$ and $\bar{\vartheta}^{*b}_n$ similarly for $b=1,\ldots,B$. Then we take $\hat\Sigma$ as the sample covariance matrix of $\big(\sqrt{n}(\hat{\vartheta}^{b}_n-\vartheta_0)^{\text T},\sqrt{n}(\hat{\vartheta}^{*b}_n-\bar{\vartheta}^{*b}_n)^{\text T}\big)^{\text T}$. The asymptotic covariance matrix of $\sqrt{n}(\bar{\vartheta}_n-\vartheta_0)$ can be estimated by $\hat\Sigma_{11}-\hat\Sigma_{12}\hat\Sigma_{22}^{-1}\hat\Sigma_{21}$.

\section{Asymptotic Properties}
Let $\theta_0=(\vartheta_0,\Lambda_0)$ denote the true value of $\theta=(\vartheta,\Lambda)$ in model \eqref{model}. Also, let $\theta^*_0=(\vartheta^*_0,\Lambda^*_0)$ be the value of $\theta^*=(\vartheta^*,\Lambda^*)$ that minimizes the Kullback-Leibler divergence given by 
$$
KL(\theta^*) = E\left\{\log\left(\frac{L(O^*)}{L(\theta^*\,|\, O^*)}\right)\right\},
$$
where $L(\theta^*\,|\, O^*)$ denotes the likelihood function at $\theta^*$ under the working model \eqref{model_work} based on the data $O^*=\{K,\,U_1,\ldots,U_K, \, \Delta_1,\ldots,\Delta_K,\,Z,\,X^*\}$, and $L(O^*)$ denotes the true likelihood of $O^*$. We first describe the regularity conditions needed as follows:
\begin{enumerate}
	\item [(C1)] $\vartheta_0$ is an interior point of a compact set $\mathcal{B}$ in $R^d$. $\Lambda_0(\cdot)$ is continuously differentiable up to order $r$ with strictly positive derivative $\lambda_0(\cdot)$ on $[\sigma,\tau]$ and $0<\Lambda_0(\sigma)<\Lambda_0(\tau)<\infty$, where $[\sigma,\tau]$ is the union of the supports of $\{U_k: k=1,\ldots,K\}$ with $0<\sigma<\tau<\infty$.
	\item [(C2)] Let $W=(X^{\text T},Z^{\text T})^{\text T}$. The distribution of $W$ has a bounded support in $R^d$. If $a^{\text T}W+b=0$, then $a=0$ and $b=0$. For some $\kappa>0$, $a^{\text T} \,\text{var}(W\,|\,K,\mathcal{U}) \,a \geq \kappa \, a^{\text T} \,E(W W^{\text T}\,|\, K,\mathcal{U}) \,a$ for all $a\in R^d$, where $\mathcal{U}=(U_1,\ldots,U_K)$.
	\item [(C3)] $\theta^*_0$ is the unique value of $\theta^*$ that minimizes $KL(\theta^*)$, and $\vartheta^*_0$ is an interior point of a compact set $\mathcal{B}^*$ in $R^{d^*}$. $\Lambda^*_0(\cdot)$ is continuously differentiable up to order $r$ with strictly positive derivative $\lambda^*_0(\cdot)$ and $0<\Lambda^*_0(\sigma)<\Lambda^*_0(\tau)<\infty$.
	\item [(C4)] Let $W^*=(X^{*{\text T}},Z^{\text T})^{\text T}$. The distribution of $W^*$ has a bounded support in $R^{d^*}$. If $a^{\text T}W^*+b=0$, then $a=0$ and $b=0$. For some $\kappa^*>0$, $a^{\text T} \,\text{var}(W^*\,|\, K,\mathcal{U}) \,a \geq \kappa^* \, a^{\text T} \,E(W^* W^{* \text T}\,|\, K,\mathcal{U}) \,a$ for all $a\in R^{d^*}$, where $\mathcal{U}=(U_1,\ldots,U_K)$.
	\item [(C5)] The number of examination times $K$ is positive with $E(K)<\infty$. There exists some constant $c>0$ such that $Pr(\min_{0\leq k\leq K}(U_{k+1}-U_k)\geq c \mid K, Z) =1$.
	The conditional densities of $(U_k,U_{k+1})$ given $K$ and $(X,X^*,Z)$, denoted by $g_k(u,v)$ for $k=0,\ldots,K$, have continuous second-order partial derivatives with respect to $u$ and $v$ when $v-u\geq c$, and are continuously differentiable with respect to $(X,X^*,Z)$.
	\item [(C6)] The degree of Bernstein polynomials satisfies $m = o(n^\nu)$ with $1/(2r)<\nu<1/2$, and $M_n = O(n^a)$ with $a>0$ controlling the size of the sieve space $\Theta_n$.
\end{enumerate}

\noindent
{\bf Theorem 1.} {\it Under Conditions (C1)--(C6), we have 
	$$
	\sqrt{n}(\bar{\vartheta}_n-\vartheta_0) = \sqrt{n}(\hat{\vartheta}_n-\vartheta_0) -\Sigma_{12}\Sigma_{22}^{-1}\sqrt{n}(\hat{\vartheta}^*_n-\bar{\vartheta}^*_n) + o_p(1) \rightarrow N(0,\Psi)
	$$
	in distribution with
	$\Psi = \Sigma_{11}-\Sigma_{12}\Sigma_{22}^{-1}\Sigma_{21},$
	where 
	$$
	\Sigma_{11} = I(\vartheta_0)^{-1} E\left\{\frac{1}{\pi_q(\Delta)}\, \big[l(\vartheta_0,\Lambda_0; O)\big]^{\otimes 2}\right\} I(\vartheta_0)^{-1} ,
	$$
	$$
	\Sigma_{22} = I^*(\vartheta^*_0)^{-1} E\left\{ \frac{1-\pi_q(\Delta)}{\pi_q(\Delta)} \big[l^*(\vartheta^*_0,\Lambda^*_0; O^*)\big]^{\otimes 2}\right\} I^*(\vartheta^*_0)^{-1},
	$$
	$$
	\Sigma_{12} = \Sigma_{21}^{\text T} = I(\vartheta_0)^{-1} E\left\{ \frac{1-\pi_q(\Delta)}{\pi_q(\Delta)} \, l(\vartheta_0,\Lambda_0; O) \, l^*(\vartheta^*_0,\Lambda^*_0; O^*)^{\text T} \right\} I^*(\vartheta^*_0)^{-1},
	$$
	and $l(\vartheta_0,\Lambda_0; O)$, $I(\vartheta_0)$, $l^*(\vartheta^*_0,\Lambda^*_0; O^*)$ and $I^*(\vartheta^*_0)$ are defined in the Appendix.
}

\vspace{.1in}

The proof of this theorem is sketched in the Appendix. Note that there is no closed-form expression for $\Sigma$, and we estimate it using the weighted bootstrap procedure described in Section 4.

\section{Simulation Studies}
In this section, we conduct simulation studies to evaluate the finite-sample performance of our proposed method. Assume that the failure time $T$ follows the Cox model 
$$
\Lambda(t\,|\,X,Z)=\Lambda(t) \exp(\beta X+\gamma Z),
$$
where $\Lambda(t)=0.2t^2$, $\beta=0$ or $0.3$, and $\gamma=0.5$. Consider two setups for the covariates: (i) $X\sim N(0,1)$ and $Z$ is empty; and (ii) $(X,Z)$ follow a bivariate normal distribution with zero mean and covariance matrix $[1,0.2; 0.2,1]$. Generate the auxiliary variable $X^*=X+e$, where $e\sim N(0,\sigma^2)$ and $\sigma=0.30$, $0.86$, or $1.70$, such that the correlation between $X$ and $X^*$ is $\rho=0.95$, $0.75$, or $0.50$, respectively. To generate the examination times, we first define a set of scheduled examination times, $s_j=ju/(n_t+1)$ for $j=1,\ldots,n_t$, where $n_t$ is the total number of scheduled examinations, and $u$ is the end-of-study time. For the $i$th subject, the actual set of examination times are $\{s_j+\epsilon_{ij}:R_{ij}=1,\,j=1,\ldots,n_t\}$, where $\epsilon_{ij}$'s are i.i.d. $\mathrm{Unif}(-t_d/3,t_d/3)$ variables, $R_{ij}$'s are i.i.d. $\mathrm{Bernoulli}(0.8)$ variables, and $t_d=u/(n_t+1)$. This is to mimic an actual follow-up study, where subjects may miss a scheduled visit and may also visit at a time different from the scheduled time. The number of scheduled examinations $n_t$ is taken as $12$. The value of $u$ is chosen such that the case rate is $p_c=0.1$, $0.2$, or $0.3$. We consider a case-cohort study for $p_c=0.1$ or $0.2$ by taking all cases, and also consider a generalized case-cohort study for $p_c=0.2$ or $0.3$ by taking a subsample of cases outside the subcohort with the sampling probability equal to $q_c=0.5$. The subcohort is selected via independent Bernoulli sampling with success probability $q_s=0.2$. The weighted bootstrap procedure for variance estimation is based on $500$ samples. The sample size is $n=1000$. The results are based on $1000$ replicates. 

Tables~\ref{simul_xonly}--\ref{simul_xz} present the estimation results of the Euclidean parameters, including ``Bias": the average point estimate minus the true parameter value, ``SSD": the sample standard deviation of point estimates, ``ESE": the estimated standard error based on weighted bootstrap, and ``CP": the coverage proportion of the 95$\%$ confidence interval based on normal approximation. The original IPW estimator of \cite{zhou2017case}, denoted by ZZC, is included for comparison with the proposed update estimator. The relative efficiency of the proposed estimator with respect to ZZC, denoted by ``RE", is given in Tables~\ref{simul_xonly}--\ref{simul_xz}. We set the degree of Bernstein polynomials to be $m=3$ for both methods. Based on our numerical experiences, the results are largely insensitive to the choice of $m$. For all setups considered, the proposed estimator is virtually unbiased, the estimated variance based on weighted bootstrap reflects the true variability, and the coverage proportion is close to the nominal level. In addition, the proposed estimator is more efficient than the ZZC estimator, and its efficiency gain for the estimation of $\beta$ increases with the case rate $p_c$ and with the association between $X$ and $X^*$. Comparing the results from the case-cohort study and the generalized case-cohort study when $p_c=0.2$, it seems that the proposed estimator gains more efficiency over ZZC under the generalized case-cohort study, which could be due to the presence of more out-of-sample information that we can borrow via the update approach.

\begin{table}
  \renewcommand*{\arraystretch}{0.85}
	\caption{Simulation results with $X$ only}\smallskip
	\label{simul_xonly}
	\resizebox{\linewidth}{!}{
	\begin{tabular}{ccccr ccccc rcccc}
		\hline
		      &       &          &        &     \multicolumn{5}{c}{$\beta=0$}      &  &    \multicolumn{5}{c}{$\beta=0.3$}     \\ \cline{5-9}\cline{11-15}
		$p_c$ & $q_c$ &  Method  & $\rho$ &     Bias & SSD   &  ESE  &  CP  &  RE  &  & Bias     &  SSD  &  ESE  &  CP  &  RE  \\ \hline
  0.10  &   1   &   ZZC    && $-$0.000 &  0.126 &  0.125 &   0.95 &   1.00 && $-$0.006 &  0.141 &  0.133 &   0.95 &   1.00 \\
 &       & Proposed & $0.95$ & $-$0.001 &  0.117 &  0.116 &   0.95 &   1.17 && $-$0.010 &  0.129 &  0.123 &   0.95 &   1.20 \\
 &       &          & $0.75$& $-$0.000 &  0.121 &  0.120 &   0.95 &   1.09 && $-$0.007 &  0.134 &  0.127 &   0.95 &   1.10 \\
 &       &          & $0.50$&  0.000 &  0.122 &  0.122 &   0.95 &   1.07 && $-$0.006 &  0.137 &  0.130 &   0.95 &   1.06 \\
0.20  &   1   &   ZZC    &&  0.009 &  0.108 &  0.103 &   0.95 &   1.00 &&  0.001 &  0.112 &  0.106 &   0.94 &   1.00 \\
 &       & Proposed & $0.95$ &  0.008 &  0.095 &  0.091 &   0.95 &   1.30 && $-$0.003 &  0.098 &  0.094 &   0.94 &   1.32 \\
 &       &          & $0.75$&  0.007 &  0.100 &  0.094 &   0.95 &   1.17 && $-$0.001 &  0.103 &  0.099 &   0.94 &   1.20 \\
 &       &          & $0.50$&  0.008 &  0.104 &  0.099 &   0.94 &   1.08 &&  0.001 &  0.108 &  0.102 &   0.94 &   1.09 \\
0.20  &   0.5   &   ZZC    &&  0.012 &  0.113 &  0.115 &   0.96 &   1.00 &&  0.006 &  0.120 &  0.118 &   0.95 &   1.00 \\
 &       & Proposed & $0.95$ &  0.006 &  0.092 &  0.095 &   0.96 &   1.49 &&  0.001 &  0.102 &  0.098 &   0.94 &   1.38 \\
 &       &          & $0.75$&  0.008 &  0.100 &  0.101 &   0.95 &   1.26 &&  0.003 &  0.109 &  0.106 &   0.96 &   1.22 \\
 &       &          & $0.50$&  0.010 &  0.106 &  0.109 &   0.95 &   1.12 &&  0.005 &  0.114 &  0.112 &   0.95 &   1.11 \\
0.30  &   0.5   &   ZZC    &&  0.006 &  0.097 &  0.098 &   0.95 &   1.00 &&  0.009 &  0.101 &  0.099 &   0.94 &   1.00 \\
 &       & Proposed & $0.95$ &  0.003 &  0.072 &  0.073 &   0.95 &   1.81 &&  0.004 &  0.077 &  0.075 &   0.96 &   1.70 \\
 &       &          & $0.75$&  0.005 &  0.082 &  0.084 &   0.95 &   1.39 &&  0.008 &  0.088 &  0.085 &   0.95 &   1.32 \\
 &       &          & $0.50$&  0.005 &  0.092 &  0.092 &   0.95 &   1.12 &&  0.009 &  0.096 &  0.093 &   0.95 &   1.11 \\ \hline
	\end{tabular}}
	\begin{tablenotes}
		\footnotesize
		\item Bias, average estimate minus true value; SSD, sample standard deviation; ESE, estimated standard error; CP, coverage proportion with 95\% nominal level; RE, relative efficiency of the proposed estimator with respect to the ZZC estimator.
	\end{tablenotes}
\end{table}

\begin{table}
  \renewcommand*{\arraystretch}{0.85}
	\caption{Simulation results with $X$ and $Z$}\smallskip
	\label{simul_xz}
	\resizebox{\linewidth}{!}{
	\begin{tabular}{ccccr ccccc rcccc}
		\hline
		      &       &          &        &     \multicolumn{5}{c}{$\beta=0$}      &  &    \multicolumn{5}{c}{$\gamma=0.5$}    \\ \cline{5-9}\cline{11-15}
		$p_c$ & $q_c$ &  Method  & $\rho$ &     Bias & SSD   &  ESE  &  CP  &  RE  &  & Bias     &  SSD  &  ESE  &  CP  &  RE  \\ \hline
		0.10  &   1   &   ZZC    & & $-$0.001 &  0.167 &  0.150 &   0.96 &   1.00 && $-$0.004 &  0.183 &  0.167 &   0.95 &   1.00 \\
 &       & Proposed &  $0.95$& $-$0.001 &  0.158 &  0.142 &   0.95 &   1.12 && $-$0.012 &  0.178 &  0.162 &   0.94 &   1.06 \\
&       &          &  $0.75$& $-$0.001 &  0.162 &  0.145 &   0.96 &   1.07 && $-$0.011 &  0.179 &  0.161 &   0.94 &   1.05 \\
 &       &          &  $0.50$ & $-$0.001 &  0.166 &  0.147 &   0.96 &   1.02 && $-$0.012 &  0.177 &  0.162 &   0.94 &   1.06 \\
0.20  &   1   &   ZZC    & &  0.000 &  0.127 &  0.120 &   0.96 &   1.00 && $-$0.012 &  0.142 &  0.134 &   0.96 &   1.00 \\
 &       & Proposed &  $0.95$& $-$0.001 &  0.117 &  0.111 &   0.96 &   1.18 && $-$0.016 &  0.137 &  0.129 &   0.96 &   1.08 \\
&       &          &  $0.75$&  0.001 &  0.119 &  0.113 &   0.96 &   1.13 && $-$0.016 &  0.138 &  0.128 &   0.95 &   1.07 \\
 &       &          &  $0.50$ & $-$0.001 &  0.124 &  0.116 &   0.96 &   1.03 && $-$0.015 &  0.140 &  0.128 &   0.95 &   1.04 \\
0.20  &   0.5   &   ZZC    & &  0.002 &  0.130 &  0.131 &   0.96 &   1.00 && $-$0.018 &  0.143 &  0.143 &   0.95 &   1.00 \\
 &       & Proposed &  $0.95$& $-$0.000 &  0.113 &  0.116 &   0.96 &   1.31 && $-$0.021 &  0.137 &  0.134 &   0.95 &   1.08 \\
&       &          &  $0.75$&  0.002 &  0.118 &  0.120 &   0.97 &   1.22 && $-$0.021 &  0.135 &  0.134 &   0.94 &   1.12 \\
 &       &          &  $0.50$ &  0.001 &  0.125 &  0.126 &   0.95 &   1.08 && $-$0.021 &  0.139 &  0.133 &   0.94 &   1.06 \\
0.30  &   0.5   &   ZZC    & &  0.004 &  0.112 &  0.112 &   0.96 &   1.00 && $-$0.011 &  0.115 &  0.118 &   0.95 &   1.00 \\
 &       & Proposed &  $0.95$&  0.002 &  0.092 &  0.096 &   0.96 &   1.47 && $-$0.010 &  0.104 &  0.105 &   0.96 &   1.24 \\
&       &          &  $0.75$&  0.004 &  0.099 &  0.101 &   0.96 &   1.29 && $-$0.013 &  0.106 &  0.105 &   0.95 &   1.19 \\
 &       &          &  $0.50$ &  0.005 &  0.106 &  0.107 &   0.96 &   1.12 && $-$0.012 &  0.105 &  0.105 &   0.95 &   1.20 \\ \hline
		      &       &          &          &   \multicolumn{5}{c}{$\beta=0.3$}    &  &   \multicolumn{5}{c}{$\gamma=0.5$}   \\ \cline{5-9}\cline{11-15}
		$p_c$ & $q_c$ &  Method  & $\rho$ &  Bias  & SSD   &  ESE  &  CP  &  RE  &  & Bias   &  SSD  &  ESE  &  CP  &  RE  \\ \hline
0.10  &   1   &   ZZC    && $-$0.012 &  0.156 &  0.153 &   0.96 &   1.00 && $-$0.004 &  0.167 &  0.167 &   0.95 &   1.00 \\
 &       & Proposed &  $0.95$ & $-$0.018 &  0.148 &  0.147 &   0.95 &   1.12 && $-$0.012 &  0.162 &  0.162 &   0.95 &   1.05 \\
  &       &          &  $0.75$ & $-$0.015 &  0.151 &  0.149 &   0.96 &   1.06 && $-$0.011 &  0.163 &  0.162 &   0.95 &   1.05 \\
 &       &          &  $0.50$ & $-$0.014 &  0.154 &  0.151 &   0.96 &   1.03 && $-$0.011 &  0.164 &  0.162 &   0.95 &   1.04 \\
0.20  &   1   &   ZZC    && $-$0.011 &  0.128 &  0.122 &   0.94 &   1.00 && $-$0.015 &  0.139 &  0.135 &   0.95 &   1.00 \\
 &       & Proposed &  $0.95$ & $-$0.010 &  0.120 &  0.114 &   0.94 &   1.14 && $-$0.018 &  0.134 &  0.129 &   0.95 &   1.07 \\
  &       &          &  $0.75$ & $-$0.010 &  0.123 &  0.116 &   0.94 &   1.08 && $-$0.017 &  0.134 &  0.129 &   0.95 &   1.08 \\
 &       &          &  $0.50$ & $-$0.011 &  0.126 &  0.119 &   0.94 &   1.04 && $-$0.019 &  0.135 &  0.129 &   0.95 &   1.06 \\
0.20  &   0.5   &   ZZC    && $-$0.012 &  0.132 &  0.132 &   0.95 &   1.00 && $-$0.012 &  0.138 &  0.144 &   0.95 &   1.00 \\
 &       & Proposed &  $0.95$ & $-$0.011 &  0.120 &  0.118 &   0.96 &   1.21 && $-$0.015 &  0.129 &  0.134 &   0.95 &   1.14 \\
  &       &          &  $0.75$ & $-$0.013 &  0.123 &  0.123 &   0.95 &   1.16 && $-$0.013 &  0.133 &  0.134 &   0.95 &   1.09 \\
 &       &          &  $0.50$ & $-$0.011 &  0.128 &  0.127 &   0.96 &   1.06 && $-$0.016 &  0.131 &  0.135 &   0.94 &   1.11 \\
0.30  &   0.5   &   ZZC    && $-$0.006 &  0.114 &  0.109 &   0.95 &   1.00 && $-$0.008 &  0.119 &  0.118 &   0.95 &   1.00 \\
 &       & Proposed &  $0.95$ & $-$0.002 &  0.096 &  0.092 &   0.94 &   1.39 && $-$0.010 &  0.109 &  0.105 &   0.95 &   1.20 \\
  &       &          &  $0.75$ & $-$0.003 &  0.102 &  0.099 &   0.94 &   1.26 && $-$0.010 &  0.110 &  0.107 &   0.95 &   1.18 \\
 &       &          &  $0.50$ & $-$0.005 &  0.108 &  0.104 &   0.94 &   1.11 && $-$0.007 &  0.111 &  0.107 &   0.94 &   1.16 \\ \hline
	\end{tabular}}
	\begin{tablenotes}
		\footnotesize
		\item Bias, average estimate minus true value; SSD, sample standard deviation; ESE, estimated standard error; CP, coverage proportion with 95\% nominal level; RE, relative efficiency of the proposed estimator with respect to the ZZC estimator.
	\end{tablenotes}
\end{table}

\section{Application to the ARIC Study}
We illustrate the proposed method using a dataset on diabetes from the Atherosclerosis Risk in Communities (ARIC) study. The ARIC study is a longitudinal epidemiological observational study that began in 1987. It consists of men and women aged 45--64 at baseline recruited from four U.S. field centers. The participants received an extensive examination at baseline, including medical, social and demographic data, and then were scheduled to be re-examined on average of every three years. For each participant, the occurrence of an asymptomatic disease such as the onset of diabetes can only be observed to fall between two consecutive examinations, and thus only interval-censored failure time data were available. As with \cite{zhou2017case}, we consider assessing the effect of high-density lipoprotein (HDL) cholesterol level on the risk of diabetes in white women younger than 55 years based on the interval-censored case-cohort data. The study cohort consists of $2799$ white women younger than 55 years and $202$ have developed diabetes during follow-up. To illustrate the proposed method, we artificially create a subcohort of subjects who exclusively have measured HDL cholesterol level. The subcohort is selected via Bernoulli sampling with the success probability $q_s=0.1$ and it includes $272$ subjects. Under the case-cohort design, we supplement the subcohort with all the remaining cases (i.e., $q_c=1$). Then the case-cohort sample consists of $451$ subjects. We consider the following Cox model,
$$
\Lambda(t\,|\,X,Z) = \Lambda(t) \exp(\beta X+\gamma^{\text T} Z), 
$$
where the expensive covariate $X$ is HDL cholesterol level, and the cheap covariates $Z$ include total cholesterol level, body mass index, age, smoking status, and indicators for field centers where Center M was set as the reference class. We compare the original IPW estimator from \cite{zhou2017case}, denoted by ZZC, with the update estimator using our proposed method. We choose the degree of Bernstein polynomials as $m=3$ for both methods. The results in Table~\ref{realdata} show that the proposed estimator yields smaller standard errors and more significant results than the ZZC estimator. In particular, both methods suggest that higher HDL cholesterol level, lower total cholesterol level and lower body mass index are significantly associated with lower risk of diabetes.

\begin{table}\small
\centering
  \renewcommand*{\arraystretch}{0.85}
 \renewcommand*{\tabcolsep}{9bp}
	\caption{Analysis results for diabetes data from the ARIC study}\smallskip
	\label{realdata}	\resizebox{\linewidth}{!}{
	\begin{tabular}{ccrcc crcc}
		\hline
		                  &  &            \multicolumn{3}{c}{ZZC method}            &  &         \multicolumn{3}{c}{Proposed method}          \\ \cline{3-5}\cline{7-9}
		    Variable      &  & \multicolumn{1}{c}{$\hat{\beta}$} &   SE   & P-value &  & \multicolumn{1}{c}{$\hat{\beta}$} &   SE   & P-value \\ \hline
		 HDL Cholesterol  &  &                         $-$0.0276 & 0.0055 & 0.0000  &  &                         $-$0.0268 & 0.0054 & 0.0000  \\
		Total Cholesterol &  &                            0.0047 & 0.0022 & 0.0326  &  &                            0.0044 & 0.0017 & 0.0084  \\
		 Body Mass Index  &  &                            0.1151 & 0.0240 & 0.0000  &  &                            0.1153 & 0.0234 & 0.0000  \\
		 Current Smoking  &  &                         $-$0.3054 & 0.2947 & 0.3001  &  &                         $-$0.3042 & 0.2892 & 0.2929  \\
		       Age        &  &                            0.0059 & 0.0571 & 0.9183  &  &                            0.0105 & 0.0559 & 0.8505  \\
		    Center-F      &  &                         $-$0.1948 & 0.2570 & 0.4483  &  &                         $-$0.2116 & 0.2528 & 0.4025  \\
		    Center-W      &  &                            0.0936 & 0.2593 & 0.7181  &  &                            0.0780 & 0.2584 & 0.7628  \\ \hline
	\end{tabular}}
\begin{tablenotes}
\footnotesize
    \item SE, estimated standard error; P-value, p-value for testing $H_0:\beta=0\,$ vs $\,H_a:\beta\neq 0$.
\end{tablenotes}
\end{table}

\section{Conclusion}
We conclude with some extensions or directions for future research. First, the proposed method can be applied to case-cohort designs where the subcohort is selected by sampling without replacement or where the sampling probability depends on covariates. Also, if the cumulative baseline hazard function $\Lambda$ is of interest, we could update $\hat\Lambda_n$ similarly as updating $\hat\vartheta_n$ to obtain a more efficient estimator, though it would entail a different and more challenging theoretical treatment as $\hat\Lambda_n$ has a rate of convergence slower than $\sqrt{n}$ and is not regular. Moreover, although we focused on the Cox model, the proposed update procedure can easily be adapted to other regression models, such as the proportional odds model and semiparametric transformation models. It can also be extended to update the IPW estimators in \cite{zhou2021semiparametric} that concerns multiple interval-censored failure times and  \cite{du2021regression} that addresses informative interval censoring. Lastly, an alternative approach to improve upon the IPW method is to employ the augmented inverse probability weighting (AIPW) method given by \cite{robins1994estimation}. Although the AIPW method has been applied in various settings, it is difficult to implement in our case, as there is not a simple estimating function that can readily be used for this approach. Another alternative method is to employ maximum likelihood estimation. However, for this method, we need to model the conditional distribution of $X$ given $X^*$ and $Z$, for which imposing a parametric assumption could be too restrictive and using nonparametric methods such as kernel or sieve estimation usually suffer from the curse of dimensionality. Both directions warrant further investigation.

\section*{Acknowledgements}
This article was prepared using ARIC Research Materials obtained from the NHLBI Biologic Specimen and Data Repository Information Coordinating Center and does not necessarily reflect the opinions or views of the ARIC or the NHLBI. Qingning Zhou's work was partially supported by the National Science Foundation grant DMS-1916170. Kin Yau Wong's work was partially supported by the Hong Kong Research Grants Council grant PolyU153034/22P.

\appendix
\section*{Appendix --- Technical Proofs}
In this Appendix, we sketch the proof of Theorem 1. We first present the following lemmas needed for the proof. For $\theta_1=(\vartheta_1,\Lambda_1)$ and $\theta_2=(\vartheta_2,\Lambda_2)$, define the distance 
$$
d(\theta_1,\theta_2) = \big(\|\vartheta_1-\vartheta_2\|^2+\|\Lambda_1-\Lambda_2\|^2_{L_2[\sigma,\tau]}\big)^{1/2},
$$
where $\|\cdot\|$ denotes the Euclidean norm and $\|\cdot\|_{L_2[\sigma,\tau]}$ denotes the $L_2$-norm on $[\sigma,\tau]$. 

\vspace{.1in}

\noindent
{\bf Lemma 1.} {\it Under Conditions (C1), (C2), (C5) and (C6), we have $d(\hat{\theta}_n,\theta_0)\rightarrow 0$ almost surely and 
	$$
	d(\hat{\theta}_n,\theta_0)=O_p(n^{-\min\{(1-\nu)/2,\,\nu r/2\}}).
	$$
}

\noindent
{\bf Lemma 2.} {\it Under Conditions (C1), (C2), (C5) and (C6), we have 
	$$
	\sqrt{n}(\hat{\vartheta}_n-\vartheta_0) = I(\vartheta_0)^{-1} \left\{\frac{1}{\sqrt{n}} \sum_{i=1}^n w_i l(\vartheta_0,\Lambda_0; O_i)\right\} + o_p(1) \rightarrow N(0,\Sigma_{11})
	$$
	in distribution, with
	$$
	\Sigma_{11} = I(\vartheta_0)^{-1} E\left\{\frac{1}{\pi_q(\Delta)}\,\big[l(\vartheta_0,\Lambda_0; O)\big]^{\otimes 2}\right\} I(\vartheta_0)^{-1}
	$$
	where $v^{\otimes 2}=vv^T$ for a vector $v$, $l(\vartheta_0,\Lambda_0; O)$ and $I(\vartheta_0)$ are the efficient score and information for $\vartheta$, respectively, based on the complete data $O=\{K,\,U_1,\ldots,U_K, \, \Delta_1,\ldots,\Delta_K,\,Z,\,X\}$, and $\pi_q(\Delta) = \left(1-\sum_{k=1}^{K_i} \Delta_{ik}\right) q_s + \left(\sum_{k=1}^{K_i} \Delta_{ik}\right) \{q_s+(1-q_s)q_c\}$.
}

\vspace{.1in}

\noindent
{\bf Lemma 3.} {\it Under Conditions (C3)--(C6), we have $d(\hat{\theta}^*_n,\theta^*_0)\rightarrow 0$ almost surely and 
	$$
	d(\hat{\theta}^*_n,\theta^*_0)=O_p(n^{-\min\{(1-\nu)/2,\,\nu r/2\}}).
	$$
}

\noindent
{\bf Lemma 4.} {\it Under Conditions (C3)--(C6), we have 
	$$
	\sqrt{n}(\hat{\vartheta}^*_n-\vartheta^*_0) = I^*(\vartheta^*_0)^{-1} \left\{\frac{1}{\sqrt{n}} \sum_{i=1}^n w_i l^*(\vartheta^*_0,\Lambda^*_0; O^*_i)\right\} + o_p(1) \rightarrow N(0,\Sigma^*)
	$$
	in distribution, with
	$$
	\Sigma^* = I^*(\vartheta^*_0)^{-1} E\left\{\frac{1}{\pi_q(\Delta)} \, \big[l^*(\vartheta^*_0,\Lambda^*_0; O^*)\big]^{\otimes 2} \right\} I^*(\vartheta^*_0)^{-1} 
	$$
	where $l^*(\vartheta^*_0,\Lambda^*_0; O^*)$ is the efficient score and $I^*(\vartheta^*_0)$ is the expectation of the negative derivative of $l^*(\vartheta^*,\Lambda^*; O^*)$ with respect to $\vartheta^*$ at $(\vartheta^*_0,\Lambda^*_0)$, 
	and $\pi_q(\Delta)$ is defined as in Lemma 2.
}

\vspace{.1in}

\noindent
{\bf Lemma 5.} {\it Under Conditions (C3)--(C6), we have $d(\bar{\theta}^*_n,\theta^*_0)\rightarrow 0$ almost surely and 
	$$
	d(\bar{\theta}^*_n,\theta^*_0)=O_p(n^{-\min\{(1-\nu)/2,\,\nu r/2\}}).
	$$
}

\noindent
{\bf Lemma 6.} {\it Under Conditions (C3)--(C6), we have 
	$$
	\sqrt{n}(\bar{\vartheta}^*_n-\vartheta^*_0) = I^*(\vartheta^*_0)^{-1} \left\{\frac{1}{\sqrt{n}} \sum_{i=1}^n l^*(\vartheta^*_0,\Lambda^*_0; O^*_i)\right\} + o_p(1) \rightarrow N(0,\bar{\Sigma}^*)
	$$
	in distribution, with
	$$
	\bar{\Sigma}^* = I^*(\vartheta^*_0)^{-1} E\left\{ l^*(\vartheta^*_0,\Lambda^*_0; O^*)^{\otimes 2}\right\} I^*(\vartheta^*_0)^{-1},
	$$
	where $l^*(\vartheta^*_0,\Lambda^*_0; O^*)$ and $I^*(\vartheta^*_0)$ are defined in Lemma 4.
}

\vspace{.1in}

\noindent
{\bf Lemma 7.} {\it Under Conditions (C3)--(C6), we have 
	$$
	\sqrt{n}(\hat{\vartheta}^*_n-\bar{\vartheta}^*_n) = I^*(\vartheta^*_0)^{-1} \left\{\frac{1}{\sqrt{n}} \sum_{i=1}^n (w_i-1) l^*(\vartheta^*_0,\Lambda^*_0; O^*_i)\right\} + o_p(1) \rightarrow N(0,\Sigma_{22})
	$$
	in distribution, with
	$$
	\Sigma_{22} = I^*(\vartheta^*_0)^{-1} E\left\{ \frac{1-\pi_q(\Delta)}{\pi_q(\Delta)} \big[l^*(\vartheta^*_0,\Lambda^*_0; O^*)\big]^{\otimes 2}\right\} I^*(\vartheta^*_0)^{-1},
	$$
	where $l^*(\vartheta^*_0,\Lambda^*_0; O^*)$, $I^*(\vartheta^*_0)$ and $\pi_q(\Delta)$ are defined in Lemma 4.
}

\vspace{.1in}

Lemmas 1 and 2 have been established by  \cite{zhou2017case} in their Theorems 1 and 2. We only need to prove Lemmas 5 and 6 under possible misspecification of the working model \eqref{model_work}. Then Lemmas 3 and 4 can easily be shown as in \cite{zhou2017case}, Lemma 7 follows from Lemmas 4 and 6, and Theorem 1 follows from Lemmas 2 and 7 by Slutsky's Theorem. In the following, we sketch the proofs of Lemmas 5 and 6. 

Let $\mathbb{P}$ and $\mathbb{P}_n$ denote the true and empirical measures, respectively. Let $l(\theta^*\mid O^*)$ be the log-likelihood function under the working model \eqref{model_work} based on a single observation $O^*=\{K,\,U_1,\ldots,U_K, \, \Delta_1,\ldots,\Delta_K,\,Z,\,X^*\}$,  
\begin{equation}\label{lik_work}
	\begin{split}
		l(\theta^*\mid O^*)= \sum_{k=1}^{K+1} \Delta_{k} & \Big\{ \log\big[\exp\left\{-\Lambda^*(U_{k-1})\exp(\beta^{* \text T} X^*+\gamma^{* \text T} Z)\right\} \\
		&\quad\quad\quad - \exp\left\{-\Lambda^*(U_{k})\exp(\beta^{* \text T} X^*+\gamma^{* \text T} Z)\right\} \big]  \Big\}.
	\end{split}
\end{equation}
Let $\Theta_n^* = \mathcal{B}^* \otimes \mathcal{M}_n$, where $\mathcal{M}_n$ is defined in \eqref{sievesp}. Then $\bar{\theta}_n^*=\arg\max_{\theta^*\in\Theta_n^*} \mathbb{P}_n l(\theta^*\mid O^*)$.

Recall that $\theta^*_0=(\vartheta^*_0,\Lambda^*_0)$ is the value of $\theta^*=(\vartheta^*,\Lambda^*)$ that minimizes the Kullback-Leibler divergence given by 
$$
KL(\theta^*) = E\left\{\log\left(\frac{L(O^*)}{L(\theta^*\mid O^*)}\right)\right\},
$$
where $L(\theta^*\mid O^*)$ denotes the likelihood of $O^*$ under the working model \eqref{model_work} and $L(O^*)$ is the true likelihood of $O^*$. Then $\theta_0^*=\arg\max_{\theta^*} \mathbb{P} l(\theta^*\mid O^*)$.

To prove Lemma 5, we can first show that $\sup_{\theta^*\in\Theta_n^*}\|\mathbb{P}_n l(\theta^*\mid O^*)-\mathbb{P}l(\theta^*\mid O^*)\|\rightarrow 0$
almost surely, following \cite{zhou2017sieve}. By the definitions of $\bar{\theta}_n^*$ and $\theta^*_0$ as well as the uniqueness of $\theta^*_0$ under (C3), we can show that $d(\bar{\theta}^*_n,\theta^*_0)\rightarrow 0$ almost surely. Also, the rate of convergence $d(\bar{\theta}^*_n,\theta^*_0)=O_p(n^{-\min\{(1-\nu)/2,\,\nu r/2\}})$ can be shown by using Theorem 3.4.1 of \cite{van1996weak} similarly as in \cite{zhou2017sieve}. Furthermore, note that $\theta_0^*=\arg\max_{\theta^*} \mathbb{P} l(\theta^*\mid O^*)$, then we can establish the asymptotic normality in Lemma 6  under conditions (C3)--(C6) similarly as in \cite{wong2023semiparametric}.

\renewcommand{\baselinestretch}{1.3}
\renewcommand{\arraystretch}{1}

\end{document}